\documentclass{ws-procs9x6}            
\begin{document}
\title{A perfect-fluid spacetime for a slightly deformed mass}

\author{M. Abishev, K. Boshkayev, H. Quevedo and S. Toktarbay$^*$}

\address{Physical-Technical Faculty, Al-Farabi Kazakh National University,\\
Al Farabi av. 71, 050040 Almaty, Kazakhstan\\
 Instituto de Ciencias Nucleares, 
Universidad Nacional Aut\'onoma de M\'exico, \\
 AP 70543, M\'exico, DF 04510, Mexico\\ 
$^*$E-mail: saken.yan@yandex.com}

\begin{abstract}
We present approximate exterior and interior solutions of Einstein's equations 
which describe the gravitational field of a static deformed mass distribution. 
The deformation of the source is taken into account 
up to the first order in the quadrupole.
\end{abstract}


\bodymatter

\ 

\ 

To describe the gravitational field of a static axially symmetric mass distribution in general relativity, 
it is necessary to consider the multipole moments of the source. From a physical point of view, one expects that the quadrupole is the largest contributor 
and higher multipoles can be neglected in a first approximation. In this case, to describe the exterior field one can use, for instance,
the exact quadrupole metric ($q-$metric)\cite{quev11,zv68}. 
\begin{eqnarray}
ds^2 & & =  A^{1+q} dt^2  -A  ^{-q}\nonumber\\
&&\times\left[ \left(1+\frac{m^2\sin^2\theta}{r^2 A }\right)^{-q(2+q)} \left(\frac{dr^2}{A }+ r^2d\theta^2\right) + r^2 \sin^2\theta d\varphi^2\right] ,
\label{zv}
\end{eqnarray}
with $A=1-2m/r$, 
which has been shown to be the simplest generalization of the Schwarzschild metric containing a quadrupole parameter $q$. 
Considering the quadrupole up to the first order only, we obtain
\begin{eqnarray}
ds^2 && =  A \left( 1+ q\ln A \right) dt^2 
- r^2 \sin^2\theta \left( 1- q\ln A \right) d\varphi^2 \nonumber\\
& -&  \left[ 1 + q  \ln A - 2q \ln\left( A + \frac{m^2}{r^2}\sin^2\theta\right)\right ] 
\left(\frac{dr^2}{A } + r^2 d\theta^2\right) \ .
\label{qmatch}
\end{eqnarray}
This is an approximate solution of  Einstein's vacuum equations up to the first order in $q$. The total mass of the 
spacetime turns out to be $M_0=m(1+q)$ and the quadrupole moment is $M_2= - (2/3) q m^3$. 

The interior solution can be generated by using  the method proposed recently in Ref. \citenum{qt15}. We obtain
\begin{eqnarray}
&& ds^2  = e^{2\psi_0} (1+2\tilde q \psi_0) dt^2 - e^{-2\psi_0} (1-2\tilde q \psi_0)\nonumber\\ 
&&\times\bigg[ e^{2\gamma_0} (1+4\tilde q \gamma_0 + \tilde q \gamma_1) 
 \left( \frac{dr^2}{r^2 f^2(r)} + r^2 (\sin^2\theta - \tilde q \sin\theta\cos\theta) d\varphi^2\right)\bigg] , \label{apm}\\
&& e^{2\psi_0} = \frac{3}{2} f(R) - \frac{1}{2} f(r), \ f(r)= \sqrt{1-\frac{2\tilde m r^2}{R ^3} }, \ e^{\gamma_0} = r e^{2\psi_0} \ ,\\
&& \gamma_1 = - 2 \int\frac{1+4\pi\sin^2\theta r^2 p_0}{rf^2(r) (1+r\psi_{0,r}) \sin^2\theta + \frac{r}{ 1+r\psi_{0,r} }\cos^2\theta} dr + \kappa \ ,\\
&& \psi_{0,r} = \frac{2\tilde m r}{ R^3 f(r) [3f(R) + f(r)]}\ ,
\end{eqnarray}
where $\tilde m$, $\tilde q$, $R$ and $\kappa$ are real constants. 
This is an interior solution up to the first order in $\tilde q $ for a perfect fluid with density and pressure
\begin{eqnarray}
\rho && = \rho_0 [1+\tilde q ( 1 + \psi_0 - 4  \gamma_0 -  \gamma_1)]  ,\ \rho_0 = const. \\
p && = p_0  [1+\tilde q ( 1 + \psi_0 - 4  \gamma_0 -  \gamma_1)] ,\ 
 p_0 = \rho_0 \frac{f(r)-f(R)}{3f(R)-f(r)} \label{p0} , 
\end{eqnarray}
respectively. In the limiting case $\tilde q \rightarrow 0$, the metric (\ref{apm}) represents a perfect fluid with constant
density $\rho_0$ and pressure $p_0$ as given in Eq.(\ref{p0}). If $\tilde m = m$, this particular  solution can be matched with the exterior 
Schwarzschild metric along  a sphere of radius $R$.

In the general case $\tilde q \neq 0$, a more detailed analysis must be carried out in order to match the above approximate interior solution 
with the approximate exterior $q-$metric given in Eq.(\ref{qmatch}). First, the matching surface must be established. Then, the matching conditions must be imposed for all metric components. This would imply a relationship between the 
exterior parameters $m$ and $q$ and the interior parameters $\tilde m$, $\tilde q$, $\rho_0$ and $\kappa$. This result will be presented elsewhere.

We acknowledge the support through a Grant of the Target Program of the MES of the RK, Grant No. 1597/GF3 IPC-30,
 DGAPA-UNAM, Grant No. 113514, and Conacyt, Grant No. 166391.



\end{document}